\documentclass{aa} 
\usepackage{graphicx}
\usepackage{soul}        
\usepackage{amssymb}
\usepackage{rotating}
\usepackage{float}  

\usepackage{txfonts}
\usepackage{hyperref}
\hypersetup{
    colorlinks=true,
    linkcolor=blue,
    filecolor=blue,      
    urlcolor=blue,
    citecolor=blue,
    pdftitle={Overleaf Example},
    pdfpagemode=FullScreen,
    }
\usepackage{xspace}

\newcommand{\vlos}[1][]{$v_{los}$ }
\newcommand{\slos}[1][]{$\sigma_{los}$ }
\newcommand{\slosf}[1][]{$\sigma_{los, f}$ }
\newcommand{\slosfz}[1][]{$\sigma_{los, f\text{=0}}$ }
\newcommand{\slosfzs}[1][]{$\sigma_{los, f\text{=0.7}}$ }
\newcommand{\sloslit}[1][]{$\sigma_{los, lit}$ }

\newcommand{\logMfz}[1][]{$\log(M_\text{f=0})$}
\newcommand{\logMf}[1][]{$\log(M_\text{f})$}
\newcommand{\logMfzs}[1][]{$\log(M_\text{f=0.7})$}

\newcommand{\kms}{km\,s$^{-1}$\xspace}
\newcommand{\met}[1][]{[Fe/H]}
\newcommand{\rh}[1][]{$r_{1/2}$ }

\begin{document}

   \title{Estimating the dynamical masses of dwarf galaxies in the presence of binary-star contamination}

      \author{
    José María Arroyo-Polonio \inst{1, 2}\thanks{E-mail: jmarroyo@iac.es (IAC)},
    Giuseppina Battaglia\inst{1, 2} and  
    Guillaume F. Thomas\inst{1, 2}
    }

   \institute{Instituto de Astrofísica de Canarias, Calle Vía Láctea s/n E-38206 La Laguna, Santa Cruz de Tenerife, España.    
         \and
             Universidad de La Laguna, Avda. Astrofísico Francisco Sánchez E-38205 La Laguna, Santa Cruz de Tenerife, España.}

   \date{Submitted on December 22, 2025; received December 22, 2025; accepted February 25, 2026}

 
  \abstract
   {The line-of-sight (l.o.s.) velocity dispersion ($\sigma_{los}$) of the stellar component of ultra-faint-dwarf galaxies (UFDs) typically ranges from 3 to 6~\kms. Applying standard mass estimators, this hints at extreme dynamical mass-to-light ratios (M/L) of approximately $\sim100-5000 \text{ M}_\odot / \text{L}_\odot $ within the half-light radius ($r_{1/2}$), making UFDs the most dark matter (DM)-dominated galaxies known and critical tests for cosmological models. However, in this regime, it is a concern whether the l.o.s. velocity component of the orbital motion of undetected binary stars (binaries) is significantly inflating the observed $\sigma_{los}$ and, consequently, UFD's dynamical mass estimates.}
   {Our goal is to correct the current estimates of $\sigma_{los}$ and dynamical masses of UFDs to account for the presence of undetected binaries with single-epoch data. Additionally, we generalize our methodology to work with multi-epoch data through which a fraction of stars forming part of binary systems can be detected via velocity variations.}
   {We use the latest binary population models in the solar neighborhood to compute the expected velocity distribution of binary stars. Then, we convolve this distribution with a Gaussian to model the l.o.s. velocity distribution of UFDs in a mixture model, in which the binary fraction is a free parameter. We apply this methodology to observed UFDs whose dynamical masses are potentially inflated by binaries. In order to generalize to the multi-epoch data case, we compute the velocity distribution of undetected binaries by applying the same cuts to the models as one would apply to the observed data to remove binaries. As the datasets currently available in the literature are not suitable for this, we will only test this method using a mock dataset.}
   {We find that estimated dynamical masses of UFDs decrease by a factor of 1.5 to 3 once undetected binaries are accounted for. These corrections significantly affect considerations about DM models based on these systems. Additionally, the lower limits of the masses decrease significantly, even challenging the classification of Leo~IV, Unions~I and Sagittarius~II as galaxies. We demonstrate that a dedicated multi-epoch observational campaign spanning one year could substantially mitigate the impact of binaries, in particular if the presence of remaining undetected binaries is accounted for. Finally, we assess the expected level of binary-star contamination in DM halo density profile inferences from dynamical models of classical dwarf spheroidal galaxies, and find that it is negligible for Sculptor-like galaxies.}
   {}

   \keywords{Galaxies: dwarf -- Galaxies: Local Group --              
                Galaxies: Kinematics and dynamics -- Binaries: spectroscopic --
                Dark matter -- Methods: statistical
               }

   \authorrunning{Arroyo-Polonio J.-M. et al.}
   \titlerunning{The impact of binaries on UFDs' dynamical masses}
   \maketitle 
%
\section{Introduction}

Since the advent of digital sky surveys, an increasing number of faint dwarf galaxies (dGs) continue to be discovered in the Local Group (LG) \citep[][among others]{Willman2005, Belokurov2008, Bechtol2015, koposov2015a, Cerny2023, Smith2024}. In these systems, known as ultra-faint dwarf galaxies (UFDs)\footnote{We refer to UFDs as those LG galaxies discovered after the advent of digital sky surveys. \cite{Simon2019} defines them as dwarf galaxies with absolute magnitudes fainter than $M_\text{V}=-7.7$}, lower luminosity correlates with higher dark-matter (DM) dominance \citep[e.g.][and references therein]{Simon2019,battaglia2022b}. The faintest ones exhibit the most extreme dynamical mass-to-light ratios (M/L), reaching values of several thousands, in solar units, within the half-light radius ($r_{1/2}$), providing a unique opportunity to study the nature of DM. Key open questions include determining the minimum DM halo mass capable of hosting a galaxy, understanding how some of these systems survive in the tidal field of massive galaxies such as the Milky Way (MW) or M31 without being disrupted, and assessing whether some of them could host cored DM halos contrary to the predictions of DM-only simulations in a $\Lambda$CDM framework. UFDs are particularly intriguing regarding this last point because they are expected to host a cuspy DM halo not only in DM-only simulations but also when baryonic feedback is implemented, making them a critical testing environment for the cusp–core problem \citep{DiCintio2014,Tollet2016}.
 
Dynamical mass estimations of UFDs rely on the use of simple mass estimators. These are expressions that relate the line-of-sight velocity dispersion ($\sigma_{los}$) and the projected half-light radius $R_h$ of the stellar component of a galaxy to the 3D dynamical mass enclosed within a region where the degeneracy with the velocity anisotropy is minimal \citep[e.g. $R_h$, 1.33~$R_h$, 1.67~$R_h$, 1.77~$R_h$ and 1.8~$R_h$, ][respectively]{Walker2009a,Wolf2010, Amorisco2012, Campbell2017, Errani2018}, in this paper we specifically use the Wolf estimator for estimating dynamical masses \citep{Wolf2010}. However, the \slos estimates often face different challenges \citep[see reviews][]{Battaglia2013, Walker2013,battaglia2022b}. In particular, when dealing with faint galaxies, \slos are calculated using spectroscopic samples of only a few dozen stars, making the results highly sensitive to outliers that can artificially inflate the measured value. Additionally, UFDs' \slos are very low, typically in the range of 3–6~\kms, making them susceptible to contamination from undetected binary stars \citep{Hargreaves1996, McConnachie2010, Spencer2017}. The role of binary stars is further complicated by our lack of knowledge about their properties in UFDs. Although binaries in the solar neighborhood are often used as a reference point, studies have shown that their properties may not necessarily apply even to the larger, "classical" dwarf spheroidal galaxies. For example, \cite{Minor2013} \& \cite{ArroyoPolonio2023} find that different period distributions can describe the observed l.o.s. velocity variability in these systems for different binary fractions ($f$). The presence of binaries even complicates efforts to distinguish UFDs from stellar clusters. As fainter galaxies have been discovered, the definition of a galaxy has been revised to accommodate these systems. According to the definition of \citet{Willman2012}, a galaxy is a gravitationally bound collection of stars whose properties cannot be explained solely by baryons and Newtonian gravity. In practice, this means that systems with a significant amount of DM are classified as galaxies, while those without it are considered globular clusters. However, the presence of undetected binaries can boost the \slos values to the extent of mimicking the dynamics of DM-dominated systems. This has led to ongoing debates about the classification of certain objects. \footnote{Metallicity spread, as an indicator of prolonged chemical evolution, can also help distinguish galaxies from clusters.}

The effect of binaries on the \slos of dGs has been explored in earlier studies. In \cite{Spencer2017} the authors studied how much undetected binaries inflate the measured \slos in mock galaxies. To do so, they assumed the binary stars properties from the solar neighborhood \citep{Duquennoy1991}. Systems with lower intrinsic \slos were the most affected ones; specifically, galaxies with intrinsic \slos in the range of 0.5 to 2~\kms are observed with \slos of around 3.5 to 4~\kms for $f$ = 0.7. In \cite{McConnachie2010} the authors created mock stellar systems without DM, such that the intrinsic \slos matched the values expected for the stellar masses of several observed UFDs. Then they contaminated the mocks with binaries, assuming different $f$, and analyzed what would be the measured $\sigma_{los}$. They used the \cite{Duquennoy1991} distributions for the parameters of binary stars. They found that the probability that the observed \slos were entirely attributable to binaries given the UFD stellar mass is not negligible for some UFDs, hinting at a possible lack of DM. Beyond the LG, \cite{Dabringhausen} found binaries to be an important factor in explaining the high observed \slos of low-mass early-type galaxies formed as tidal dwarf galaxies, which are expected to contain almost no DM. However, significant progress has been made in recent years: binary star models for solar neighborhood stars have been updated \citep{Moe2017} and numerous new systems have been discovered, further blurring the distinction between dGs and stellar clusters.

Various techniques have been proposed in the literature to account for binaries when estimating \slos from multi-epoch data. Some of these methods have been applied to observed datasets, but only to a handful of systems. In \cite{Minor2010} the authors used mock data to compute the fraction of stars exhibiting significant velocity variations above a given threshold, using this fraction to provide tabulated corrections to correct the observed \slos of classical dwarf spheroidal galaxies. \cite{Martinez2011} \& \cite{Simon2011} measured the \slos of Segue 1 by accounting for binaries in a mixture model, using flexible period distributions and multi-epoch data. However, they only had repeated measurements for a third of the entire sample, with velocity errors of approximately 6~\kms. \cite{Minor2019} computed the \slos of Reticulum II, also correcting for the effect of binaries. They combined data obtained using different instruments to build a multi-epoch dataset. This makes their inference on the \slos partially sensitive to the offsets between the instruments. \cite{Buttry2022} attempted to resolve the \slos of Triangulum II by using multi-epoch data but taking a different approach. They constrained the orbital parameters of one known binary star so they could include it in the computation. Nevertheless, they could not resolve the $\sigma_{los}$. 

For the rest of the systems, the \slos has not been corrected for the effect of binary stars. It is true that most of the datasets for UFDs are single-epoch or have very short total time baselines. But binaries can still be accounted for; however, it has not yet been done in a homogeneous and systematic manner. Furthermore, while multi-epoch data is now considered essential to address these challenges, it remains unclear how much of the effect of undetected binaries can be mitigated by a given observation strategy or whether the effect of undetected binaries with longer periods will still be relevant.

To address these issues, we reexamine the impact of binary stars in inflating the observed \slos of UFDs. We present a methodology for correcting current \slos estimates from single-epoch observations similar to that presented in \cite{Martinez2011} for multi-epoch data. This methodology accounts for undetected binary stars using models based on the solar neighborhood, while making no assumptions on $f$. We apply this methodology to some UFDs with \slos measurements potentially inflated by binaries. Additionally, we generalize this methodology for multi-epoch data, accounting for both unidentified and identified binaries detected via velocity variations. This tool also provides a framework for optimizing observational strategies to robustly measure \slos in these systems.

In Sec.~\ref{sec:Met}, we describe the methodology used to correct single-epoch observations, as well as the UFDs' data used. Sec.~\ref{sec:results} presents the corrected \slos and dynamical masses for these systems. In Sec.~\ref{sec:gen_mult} we generalize the proposed approach to handle multi-epoch data, and we test it on a mock dataset. In Sec.~\ref{sec:obs_vs_int}, we revisit the relation between the observed \slos and the intrinsic \slos as a function of the binary fraction $f$, focusing on the regime of low-number statistics. Finally, in Sec.~\ref{sec:conclusions}, we summarize the main conclusions of the work.

\section{Method} \label{sec:Met}

This section describes the methodology used to correct the measured \slos in UFDs using single-epoch data for the motion of undetected binary stars. In Sec.~\ref{sec:Met_models} we present the binary star models assumed. In Sec.~\ref{sec:Met_Singleepoch} we detail the specific mixture model used to compute $\sigma_{los}$. Finally, in Sec.~\ref{sec:data} we present the datasets employed.

\subsection{Binary star models}\label{sec:Met_models}
At the distances of UFDs, binary systems are unresolved. Throughout this paper, we assume that the observed star in a binary system is the primary. The l.o.s. velocity component of the primary star in a binary system can be computed analytically \citep[e.g. see][]{GreenSphericalDynamics}: 
    \begin{equation} \label{eq:vlos_bin}
    v_{los,bin}=\frac{q\text{ }\sin(i)}{\sqrt{1-e^{2}}}\left(\frac{2\pi Gm_{1}}{P(1+q)^{2}}\right)^{1/3}\left(\cos\left(\theta+\omega\right)+e\text{ }\cos(\omega)\right).
    \end{equation}

\noindent Out of the 7 parameters that describe the motion, 4 are intrinsic to the system ($q$, $m_1$, $P$ and $e$) and 3 are extrinsic ($i$, $\omega$ and $\theta$). Among the intrinsic ones: $m_1$ is the mass of the primary star; $q=\frac{m_{2}}{m_{1}}$ is the ratio between the mass of the secondary star, $m_2$, and of the primary star; $e$ is the eccentricity of the orbit around the center of mass; and $P$ is the orbital period. Of the extrinsic parameters: $\omega$ is the argument of the periastron, the angle corresponding to the point of the orbit that is closest to the center of mass; $i$ is the inclination, i.e., the angle between the plane determined by the binary system's orbit and the plane perpendicular to the l.o.s.; and $\theta$ is the true anomaly, the only parameter that evolves with time, indicating the position of the star along the orbit. 

We adopt the parameter distributions from \cite{Moe2017}, derived for stars in the solar neighborhood, but with some simplifications: we set $m_1$ to 0.8 $M_{\odot}$, which is the typical mass of old metal-poor stars near the turn-off point or in the red giant branch (RGB)\footnote{For example, according to a synthetic population based on Basti \citep{Basti2013} stellar evolutionary models and a Salpeter initial mass function for a constant SFH between 12-13 Gyr ago and a metallicity distribution centered at [Fe/H] = -2.3 dex with a spread of 0.5 dex, the mean mass of the stars around the main-sequence turn-off and the RGB is 0.803 $M_{\odot}$.}. We assume a fixed $q$ distribution independent of the orbital period. Specifically, we adopt $\gamma_{\mathrm{small}q}$ = 0.3, $\gamma_{\mathrm{large}q}$ = -0.65 and $\mathcal{F}_ \mathrm{twin}$ = 0.15, following the parameters presented in \cite{Moe2017}. This distribution provides an approximate description across the full range of periods. For periods $\log(P [days]) < 1$, we assume circular orbits $e= 0$, while for $\log(P [days]) > 1$ we adopt a uniform $e$ distribution between 0 and 1. The small modifications applied to the $q$ and $e$ distributions are intended to improve the computational efficiency of the parameter sampling. We have verified that reasonable variations of these assumptions do not affect the results presented below.

The minimum orbital separation ($a_{\min}$) is set according to Eq. 5 of \cite{ArroyoPolonio2023} for each value of $q$ and the radius of the primary star, in order to avoid Roche-lobe overflow. We use 0.21 au as the typical radius for a 0.8 $M_{\odot}$ RGB star ($\log g[cm~s^{-2}]=1$) and 0.007 au for a main-sequence MS star close to the turn-off point ($\log g[cm~s^{-2}]=3.9$). Then, we use $a_{\min}$ to set a minimum cutoff for the period distribution using Kepler’s third law. To sample the extrinsic parameters we follow the approach described in \cite{ArroyoPolonio2023}.

In our models, we assume that all the stars belonging to a galaxy are either RGB or MS stars, according to the dominant population in the dataset. The main difference with respect to the previous models from \cite{Duquennoy1991} used in \cite{Minor2010}, \cite{McConnachie2010} \& \cite{Spencer2017}, among others, is that the period and the mass ratio distributions are specifically computed as a function of the mass of the primary star.

\subsection{Mixture model including binaries}\label{sec:Met_Singleepoch}
We will work within the Bayesian Inference framework. To account for measurement errors in \vlos and the motion of undetected binaries with a free parameter $f$ when computing the \slos of a system, we define the likelihood: 
\begin{equation} \label{eq:like}
     \begin{aligned}
    \mathcal{L} &= \prod_i^{n_{star}}\left[f \left(N_i(v_{sys}, \sigma_{los}) \ast \mathcal{B}_i\right) + (1-f) N_i(v_{sys}, \sigma_{los}) \right], \text{where} \\
    &N_i(v_{sys}, \sigma_{los}) = \frac{1}{\sqrt{2\pi(\sigma_{los}^2 + \Delta v_{los,i}^2)}}
    \exp{\left( -\frac{(v_{sys} - v_{los,i})^2}{2(\sigma_{los}^2 + \Delta v_{los,i}^2)} \right)}.
\end{aligned}
\end{equation}

\noindent In the above, $\mathcal{B}_i$ is the velocity distribution of the primary stars in a binary system for the $i$-th star (in the following, we explain why it depends on the individual observed star), the symbol $\ast$ denotes convolution, and $n_{star}$ is the number of stars in the sample. There are three free parameters: the systemic l.o.s. velocity of the galaxy $v_{sys}$, $\sigma_{los}$, and $f$. Finally, $v_{los, i}$ and $\Delta v_{los, i}$ are the heliocentric l.o.s. velocity of the $i$-th star and its associated error. We include $v_{sys}$ as a free parameter rather than using literature values, as the inclusion of binaries can also affect the recovered systemic l.o.s. velocity, mostly for very asymmetric velocity distributions. This methodology is broadly similar to the one presented in \cite{Martinez2011}, in that we both account for binaries in a mixture model. However, there are some differences in how we compute $\mathcal{B}_i$, as they also deal with contaminants from the MW, whereas we use clean samples composed only of likely member stars. \cite{Gration2025} also presented a similar mixture model formalism that incorporates not only spectroscopic binaries but also visual binaries. However, their method has not yet been applied to observed data.

$\mathcal{B}_i$ is computed from $10^7$ Monte Carlo samples\footnote{We verified that the variation in the shape of $\mathcal{B}_i$ for a number of realizations larger than $10^7$ is negligible.} drawn from the parameter distributions in Sec.~\ref{sec:Met_models} and evaluated using Eq.~\ref{eq:vlos_bin}. However, some cuts must be applied to resemble what is done in observational studies: stars in the field of the UFD but with \vlos inconsistent with the galaxy's velocity distribution (e.g., deviating by more than $3 \times \sigma_{los}$ from the systemic velocities) are not considered UFD members but rather MW contaminant stars. By applying this cut, some binary stars are also removed. Therefore, this must be reflected in our representation of $\mathcal{B}_i$; otherwise, our models could include stars with, for example, $v_{los,bin} > 50$~\kms, which are not actually included in the datasets of likely member stars ($3 \times \sigma_{los}$ is typically lower than 30~\kms for UFDs and always lower than 15~\kms for the systems studied in this work; see Tab.~\ref{tab.1}). We implement this in our models by truncating the $B_i$ distribution at $\pm3 \times \sqrt{\sigma_{los}^2 + \Delta v_{los,i} ^2 }$. This excludes stars for which $v_{los, bin}$ is already in a 3-$\sigma$ tension with respect to the UFD's velocity distribution, even for a star with an intrinsic l.o.s. velocity $v_{los, int} = v_{sys}$. In this step, we use the value of \slos computed without accounting for binaries ($\sigma_{los,f=0}$). Since the velocity errors vary for each star, different cuts must be applied to the $\mathcal{B}_i$ distribution on a star-by-star basis. We warn that, in this step, we remove binaries from both the data and the models. Therefore, the quoted binary fraction $f$ corresponds to the cleaned data set and not to the intrinsic binary fraction of the galaxy. For reference, for the case of Car~II ($\sigma_{los,f=0}$ = 3.24 \kms) and a velocity uncertainty of 1 \kms, the fraction of removed binaries is 4.6\%, which is representative of the average value across the sample.

We compute our inference within a Bayesian framework. To explore the parameter space and sample from the posterior probability distribution (PPD), we use the EMCEE package \citep{Foreman-Mackey2013}, a Python implementation of the Affine Invariant Markov Chain Monte Carlo (MCMC) ensemble sampler \citep{Goodman2010}. Further details about the MCMC runs can be found in App.~\ref{app:emcee}.

\subsection{Data}\label{sec:data}

We analyze UFDs that are either confirmed galaxies or candidate galaxies whose \slos may be inflated by undetected binary stars, specifically those with \slos < 4.5~\kms. These UFDs are more susceptible to the influence of undetected binaries. According to \cite{McConnachie2010}, binaries alone cannot produce a much larger observed \slos in a system without DM. Furthermore, \cite{Spencer2017} shows that for DM-dominated systems with around 100 tracers, only those with intrinsic \slos lower than 4~\kms can have observed \slos inflated due to binary stars by more than 50$\%$. The systems selected based on this criterion are Bootes I, Carina II, Crater II, Eridanus III, Hydrus I, Leo IV, Leo V, Reticulum II, Sagittarius II, Segue 1, UNIONS 1 / Ursa Major III, and Willman 1 (Boo~I, Car~II, Cra~II, Eri~III, Hyd~I, Leo~IV, Leo~V, Ret~II, Sag~II, Seg~1, Uni~1 and Wil~1 from now on) (for the list of original references, see the notes of Tab.~\ref{tab.1})\footnote{Aquarius III and Delve 1 are special cases not included in the analysis; see App.~\ref{app:special} for details.}. For each system, we use the most recent spectroscopic datasets available in the literature. Details on how we process the data and the relevant references are provided in App.~\ref{app:data}.

\section{Correcting UFDs \slos from single-epoch data}\label{sec:results}
In this section, we present the \slos corrected for the presence of binary stars for the UFDs introduced in Sec.~\ref{sec:data}. The quoted values and errors (Tab.~\ref{tab.1}) are presented in the following format: $\text{quantity} = \text{median}^{+1\sigma(+3\sigma)}_{-1\sigma(-3\sigma)}$, where the 1- and 3-$\sigma$ values represent the difference between the [16$^{th}$-84$^{th}$] and [0.15$^{th}$-99.85$^{th}$] percentiles and the median of the PPD of $\sigma_{los}$, respectively. For each UFD, we compute the \slos under three different assumptions: without correcting for binary stars ($f$=0), assuming a flat prior on $f$, and a fixed $f$ = 0.7 (in practice selecting 0.65 < $f$ < 0.75 on the PPD)\footnote{We adopt $f = 0.7$ as an extreme yet plausible value,  consistent with the upper bound of the $1\sigma$ range of the expected binary fraction for solar-neighbourhood systems at [Fe/H]$=-3$ \citep{Moe2019}. In addition, the only direct constraint on $f$ for a UFD comes from \cite{Minor2019}, who find $f > 0.5$ at the 90\% confidence level for Ret~II.}, labeled as $\sigma_{los, f\text{=0}}$, $\sigma_{los, f}$, and $\sigma_{los, f\text{=0.7}}$, respectively. We also compute the dynamical mass enclosed within the 3D half-light radius ($r_{1/2}$) in all the cases (labeled as \logMfz, \logMf, \logMfzs). To do so, we use the Wolf mass estimator \citep{Wolf2010}: 

\begin{equation}
    M_{dyn}(<r_{1/2}) = \frac{4}{G} R_{1/2} \sigma_ {l.o.s.}^2, \label{eq:wolf}
\end{equation}\label{eq:massestimator}

\noindent where $M_{dyn}(< r_{1/2})$ is the mass within the 3D half-light radius, $R_{1/2}$ is the projected half-light radius, and $G$ the gravitational constant. Since UFDs appear flattened on the sky, we use the circularized radius $R_{1/2}=a_{1/2}\sqrt{1-e}$, where $a_{1/2}$ is the semi-major axis of the projected half-light ellipse and $e$ is the ellipticity of the system. 

In the following, we discuss the results that apply to the overall population of UFDs, as well as some specific compelling cases. These results highlight the importance of accounting for the effect of binaries to obtain robust \slos estimates. In Sec.~\ref{sec:gen_mult} we show how multi-epoch data is essential for this purpose. In App.~\ref{app:classical}, we briefly analyze whether the effect of binaries is also important for the determination of the DM density profiles of classical dwarf spheroidals.

\begin{figure*}
    \centering
        \includegraphics[width=\textwidth]{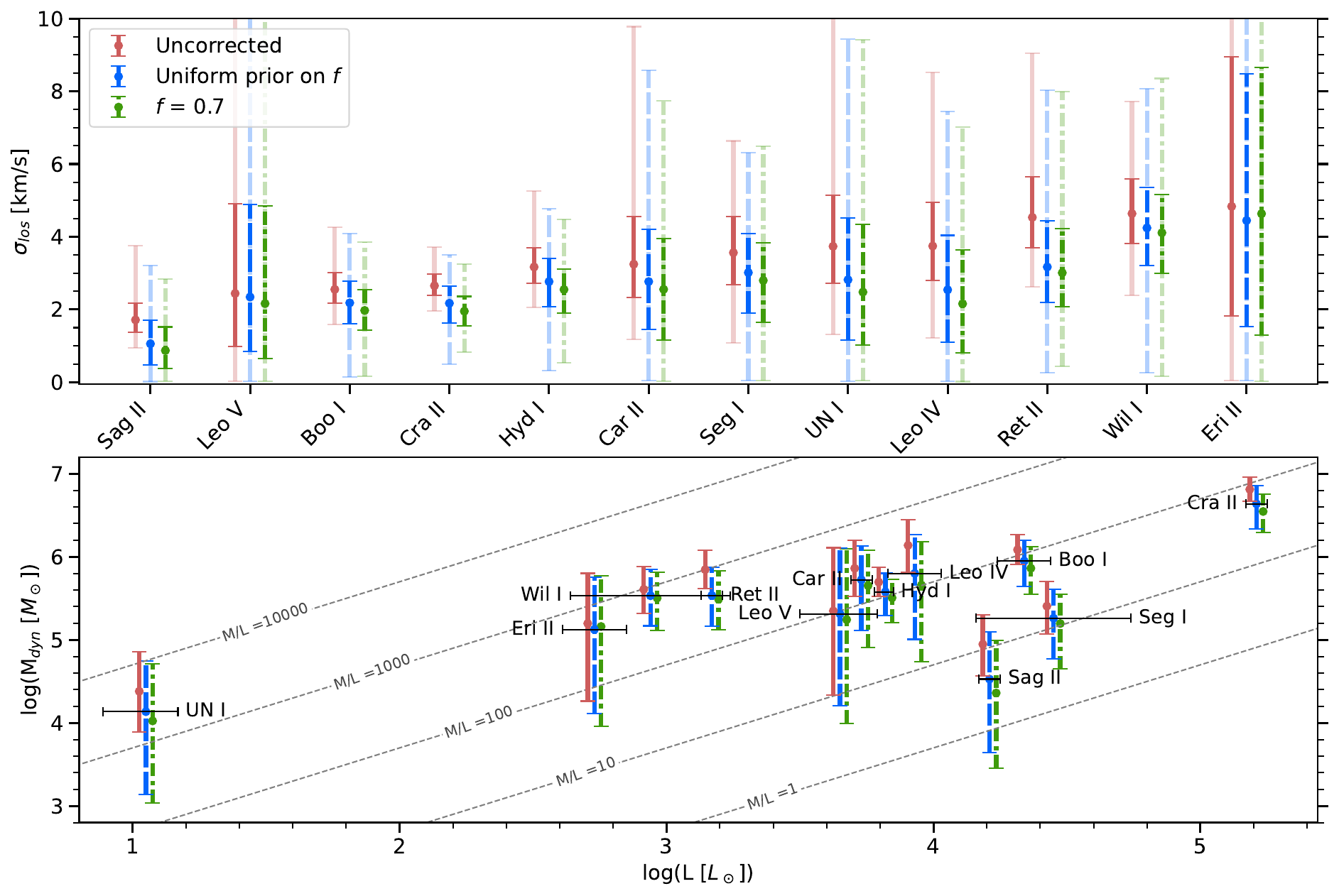}

        \caption{Observed \slos and estimated masses of UFDs. Upper panel: Observed \slos for the UFDs analyzed in this work sorted by $\sigma_{los, f\text{=0}}$. Lower panel: Mass estimated with Eq.~\ref{eq:wolf} versus the luminosity, both in log-scale. The diagonal gray dashed lines indicate regions of constant M/L (= $M_{dyn}(<r_{1/2})/(L/2)$) in solar units indicated by the numbers. The horizontal black lines show the 1$\sigma$ ranges for $L$. The red, blue, and green vertical lines indicate the median and 1-$\sigma$ ranges for the values uncorrected by binaries, corrected assuming a uniform prior on $f$ and corrected assuming $f$ = 0.7, respectively. The shaded lines in the upper panel indicate the 3-$\sigma$ range.}
        \label{fig:Fig1}
    \end{figure*}

\subsection{Population of UFDs} \label{sec:pop_UFDs}

Fig.~\ref{fig:Fig1} shows the \slos values of the analyzed UFDs and the positions of these in the $M_{dyn}(<r_{1/2})$ vs. luminosity plane (from now on, the luminosity values refer to the V-band). Undetected binaries typically inflate the median \slos\ of the analyzed UFDs by a factor of approximately 1.25-1.75. This inflation translates into differences in dynamical mass estimates (see eq.~\ref{eq:massestimator}) ranging from a factor of $\sim$ 1.5 for the least affected UFDs, such as Boo~I or Eri~III, up to a factor of $\sim$ 3 for the most affected ones, such as Leo~IV, Uni~1, Sag~II, or Seg~1. The variation in this effect among galaxies arises from differences in the observed \slos itself, as well as in the individual $\Delta v_{los,i}$ and the overall observed \vlos distribution. The correction becomes smaller when the observed distribution closely follows a Gaussian profile without extended tails that could be attributed to binary stars, or when $\Delta v_{\mathrm{los},i}$ are large compared to the intrinsic velocity scatter. Our results align with those of \citet{Spencer2017}, who predicted that a system with an intrinsic \slos = 2~\kms would be observed with \slos = 3~\kms for $f=0.7$. A similar trend is observed for some UFDs when comparing the columns \slosfzs and \slosfz of Tab.~\ref{tab.1}.

The overall decrease in the estimated dynamical masses of UFDs impacts existing considerations about the density profile of their DM halo. For instance, \citet{Errani2022} demonstrated that the relation between the circular velocity and the size of UFDs can be reasonably well explained if these systems inhabit NFW halos embedded within the MW's tidal field. However, our revised mass estimates imply lower circular velocities for UFDs, thereby exacerbating existing tensions with this scenario, as in the case of Cra~II, and revealing new ones, such as for Boo~I should it have a binary fraction $f$ > 0.5. The overall decrease in dynamical mass also has significant implications for the expected DM annihilation or decay signal in UFDs. In Tab.~\ref{tab.1}, we present the updated J-factors of the analyzed UFDs, computed using Eq. 16 from \cite{Evans2016}. They are computed using the binary-corrected dynamical mass (\logMf) and an angular aperture of 0.5$^\circ$. Although the effect on the median values is modest, the new errors render the lower limits of the J-factors highly uncertain.

Even for systems where the correction for binaries is not large, the PPDs of \slos exhibit significant tails extending towards \slos of 0~\kms when binaries are taken into account; see Fig.~\ref{fig:FigB1}. These tails are propagated into the uncertainties of the dynamical masses as well. Consequently, although the median $M_{dyn}(<r_{1/2})$ is typically higher for the fainter galaxies, almost all UFDs remain consistent with an M/L($<r_{1/2}$) ratio close to 100 $\text{M}_\odot/\text{L}_\odot$ or lower within 1$\sigma$ when binaries are accounted for, as shown in Fig.~\ref{fig:Fig1}. For the systems with the lowest mass limits, our results are consistent with those of \cite{McConnachie2010}, who found that, for some UFDs, there is a non-negligible probability that binary stars could inflate the observed \slos from M/L values typical of globular clusters. For example, they identified Seg~1 and Leo~IV among the most likely systems to exhibit this behavior. In our analysis, the 2.5th percentile of the \slosf PPD for these galaxies (corresponding to the lower limit for a 2$\sigma$ confidence interval), is 0.60~\kms for Seg~1 and 0.19~\kms for Leo~IV. There is therefore a 2.5\% probability that the true \slos of these systems lies below these values, which are comparable to the expected \slos if one assumes the typical M/L of globular clusters for these UFDs, around 0.2~\kms \citep{McConnachie2010}. We also find several recently discovered UFDs to exhibit very low lower limits on $\sigma_{los}$. Interestingly, no UFD in our sample has a \slosf exceeding 0.5~\kms within 3-$\sigma$ (see Table~\ref{tab.1}).

\subsection{Individual UFDs}\label{sec:Ind_UFDs}

The results for some UFDs are particularly noteworthy. For Leo~IV, Sag~II, and Uni~1 the velocity dispersion ($\sigma_{los, f\text{=0}}$) is resolved\footnote{That is, the PPD of \slos does not include $0~\mathrm{km~s^{-1}}$, implying that the probability of \slos being strictly greater than zero is 100\%. In the context of UFDs and given the precision on the \slos estimates, this suggests the presence of a significant DM halo.} when the impact of undetected binary stars is neglected. In contrast, when the effect of binary stars is accounted for, the l.o.s. velocity dispersion ($\sigma_{los, f}$) becomes unresolved, as shown in Fig.~\ref{fig:FigB1}. The internal kinematics of these systems are therefore consistent with a negligible DM content. Consequently, robustly distinguishing these systems from globular clusters requires additional diagnostics, such as their chemical properties.

{\bf Leo~IV: } 
Although the metallicity dispersion is formally resolved \citep{Jenkins2021}, excluding a single star with borderline metallicity renders it unresolved. Hence, the detection of a metallicity spread in this system is not robust, and additional information is required for it to be considered a confirmed UFD.

{\bf Sag~II: } 
The metallicity spread of Sag~II is currently unconstrained \citep{Longeard2021}; therefore, its classification as a dwarf galaxy cannot be confirmed based on either its internal kinematic properties or from chemical enrichment arguments. Recent analyses of other elemental abundances in Sag II suggest that the system remains consistent with both globular clusters and UFDs \citep{Zaremba2025}. This is consistent with \cite{Baumgardt2022} who reported evidence for mass segregation in this system and concluded that it is likely a globular cluster. 

{\bf Uni~1: } 
The metallicity spread of Uni~1 is also unconstrained \citep{Simon2024, Cerny2025}, and its classification as a galaxy therefore remains uncertain. Notably, for Uni~1 to survive the MW field as a bound system, it must have formed at the center of at least a 10$^9$ M$_{\odot}$ cuspy $\Lambda$CDM halo \citep{Errani2024}, consistent with the inference from the mass estimators neglecting binaries. However, this does not mean that it could not be the remnant of a dissolved globular cluster. Interestingly, recent multi-epoch observations show no evidence of DM in this system \citep{Cerny2025}. These authors report a 95\% upper limit \slos of 2.3~\kms, with a PPD similar to that obtained from our single-epoch data with models including binaries, but more tightly constrained. Thus, none of the three UFDs whose \slos become unresolved after accounting for binaries can be confirmed as galaxies from chemical analyses either.

{\bf Ret~II: } 
Ret~II is another compelling UFD, but for the opposite reason compared to the previously discussed systems: its \slos distribution remains well resolved even after accounting for binaries. This is consistent with the recent results of \cite{luna2025}, who found evidence of a double-peaked MDF, a clear signature of extended chemical evolution, not expected in globular clusters. The lack of clear mass segregation in this system is also consistent with its classification as a UFD \citep{Baumgardt2022}. The \slos of Ret~II has also been computed accounting for binaries with a similar methodology by \cite{Minor2019}. They used the same dataset as we did but combined it with data from \cite{Simon2015} and \cite{Ji2016}, thereby using multi-epoch information. However, this is a heterogeneous dataset acquired with different instruments that presents systematic offsets, as can be seen in their Fig. 1. The authors predict that $f$ is greater than 0.5 at the 90 percent confidence level and that the intrinsic velocity dispersion is \sloslit = $2.8^{+0.7}_{-1.2}$~\kms. These results agree very well with our inference using only \cite{Koposov2015} data and assuming $f$ = 0.7, \slosfzs$ = 3.01^{+1.22}_{-0.93}$~\kms. Although our PPD for $f$ favors larger values, it is not as extreme as theirs.

{\bf Cra~II: } 
Cra~II’s \slos decreases by a factor of 1.4 when comparing the $f = 0$ and $f = 0.7$ models, reaching a value of \slosfzs $= 1.95^{+0.41(+1.30)}_{-0.41(-1.12)}$ compared with \slosfz $= 2.65^{+0.32(+1.06)}_{-0.27(-0.69)}$ in the absence of any correction. Despite this significant reduction, the velocity dispersion remains resolved (see also Fig.~\ref{fig:FigB1}), indicating that Cra~II is still dominated by dark matter even after accounting for inflation due to unresolved binary stars. Interestingly, this correction brings the value of \slos into even closer agreement with the prediction of modified Newtonian dynamics (MOND) for this ultra-faint dwarf galaxy, \slos $= 2.1^{+0.9}_{-0.6}$~\kms \citep{McGaugh2016}. This MOND prediction includes the external field effect (EFE) of the Milky Way \citep{Bekenstein1984, Famaey2012, Milgrom2014}. The EFE leads to a decrease in the velocity dispersion of dwarf galaxies compared to the deep-MOND prediction, i.e. when the system is effectively isolated. Nevertheless, it is important to stress that extended tidal features have recently been observed around Cra~II \citep[e.g.][]{Coppi2024, Vivas2025} , confirming earlier predictions based on its orbital properties \citep{Fritz2018, Sanders2018, Simon2019, Ji2021,Battaglia2022, pace2022, Borukhovetskaya2022}. As such, it is possible that the large size and low velocity dispersion of this galaxy are at least partly driven by its current tidal disruption. However, early studies have shown that reproducing both the size and the velocity dispersion of Crater~II within a dark-matter framework requires a cored DM density profile. The additional decrease in velocity dispersion obtained when accounting for unresolved binary stars may therefore point toward an even more weakly concentrated, strongly cored dark-matter halo, as previously suggested by \citet{Sanders2018,Errani2022}.

{\bf Seg~1: } 
Seg~1 is a system in which the correction for binary stars is relatively modest, with a factor of 1.18 between \slosfz and \slosf. It is also the system with the largest number of line-of-sight velocity measurements. The well-resolved \slos is consistent with the absence of mass segregation in this system \citep{Baumgardt2022}, favoring its classification as a UFD. In \citet{Martinez2011} and \citet{Simon2011}, the authors accounted for the effect of binaries on the velocity dispersion of Seg~1 using the same dataset as in our work, but applying a multi-epoch correction based on individual exposures. They obtained a slightly different result, \sloslit = $3.7^{+1.4}_{-1.1}$~\kms, compared to our estimate \slosf$ = 3.01^{+1.07}_{-1.12}$~\kms, but still within 1-$\sigma$. Their methodology differs in that it is applied directly to individual exposures and treats both the mean and the width of the binary period distribution as free parameters with non-informative priors. Since the available data provide limited constraints on these parameters, their results may be influenced by the adopted prior limits. 

Based on our inference and the literature review, Leo~IV, Sag~II, and Uni~1 may be consistent with a globular cluster classification. This raises the question of whether these systems could be mass segregated and in a energy equipartition state, thereby violating the assumption made in Eq.~\ref{eq:like} that the velocity distribution of the center of mass of binary stars is the same to that of single stars.

In a system with full energy equipartition, the velocity dispersion scales as $\sigma \propto m^{-0.5}$. For the median value of our adopted $q$ distribution, binaries in our model are approximately 1.5 times more massive than single stars. Under full equipartition, this would imply a difference in velocity dispersion of a factor of 1.22. However, both numerical simulations and observational estimates for the Omega-Centauri cluster suggest a significantly lower degree of energy equipartition, consistent with $\sigma \propto m^{-0.15}$ \citep{Trenti2013}. In this case, the expected difference in velocity dispersion between binaries and single stars is reduced to a factor of 1.06. We therefore conclude that energy equipartition would not significantly affect the inferences presented in this work.

\section{Generalization to multi-epoch data}\label{sec:gen_mult}
In this section, we show how multi-epoch data with precise \vlos measurements at each epoch can help mitigate the inflation caused by binaries. In Sec.~\ref{sec:Met_Multiepoch} we generalize the methodology to work with multi-epoch data and, in Sec.~\ref{res:mock}, we test it on a mock galaxy.

\subsection{Correcting \slos from multiple epoch observations }\label{sec:Met_Multiepoch}

In the case of multi-epoch data, the likelihood function (Eq.~\ref{eq:like}) remains unchanged relative to the single-epoch case. However, $\mathcal{B}_i$ must be modified to account for the treatment of multi-epoch data. For example, one valid approach is to consider stars that exhibit velocity variations between two epochs greater than three times the measurement error of the velocity difference as likely binaries, and to exclude them from the analysis. This is the approach we adopt in the following mock test, although other criteria could also be applied \citep[see, for example,][]{Koposov2011, Walker2023}. This procedure leads to a non-uniform detection of binary stars, since those in systems with large velocity amplitudes and short orbital periods are more likely to be identified. Moreover, the threshold for detecting binaries depends on the individual $\Delta_{los,i} $ of each star. Finally, the error-weighted mean of the line-of-sight velocities from multiple observations of stars not detected as binaries is typically used, which also affects the shape of $\mathcal{B}_i$.

To account for the aforementioned steps, we compute $\mathcal{B}_i$ as follows. First, we generate $10^7$ sets of parameters for binary stars (see Sec.~\ref{sec:Met_models}) and compute their \vlos at the first epoch using Eq.~\ref{eq:vlos_bin}. We then evolve the true anomaly $\theta$ for each subsequent epoch of the specific dataset under consideration\footnote{See Sec.~3.2 of \cite{ArroyoPolonio2023} for a description of how the true anomaly evolves over time as a function of the period.}, and recompute the l.o.s. velocity for each binary star according to Eq.~\ref{eq:vlos_bin}. If the velocity variation of a given star between two epochs exceeds three times its error, we consider that star a detected binary and exclude it from the model. Once all the detected binaries are removed, we compute the mean of the \vlos of the undetected binaries across all epochs. The resulting distribution of averaged \vlos serves as $\mathcal{B}_i$, which can be used to directly apply the previous methodology to the error-averaged data. Note that now the shape of $\mathcal{B}_i$ is different for each star, as it depends on the individual $\Delta v_{los,i}$. 

\subsection{Test on a mock dataset} \label{res:mock}

To test our methodology, we created a mock dataset for a UFD containing 20 stars, with an intrinsic \slos $=2.4$~\kms and $f = 0.7$ (and a systemic $v_{sys}$ of 120~\kms \footnote{The value of $v_{sys}$ is completely arbitrary and has no effect on the analysis.}). The l.o.s. velocity errors of the stars were generated assuming a Gaussian distribution centered on 1~\kms with a width of 0.4~\kms and a floor at 0.2~\kms. This mimics what could, for example, be achieved with a high-resolution grating on a spectrograph such as FLAMES/GIRAFFE at VLT. We explore two cases: a single-epoch case and a multi-epoch case. For the multi-epoch case, we simulated observations of two additional epochs, $t_1=t_0+\text{3 months}$ and $t_2=t_0+\text{12 months}$, for a total time baseline of one year. For simplicity, we assume that $\Delta v_{los,i}$ is the same for a given star at each epoch.

\begin{figure*}
    \sidecaption
        \includegraphics[width=12cm]{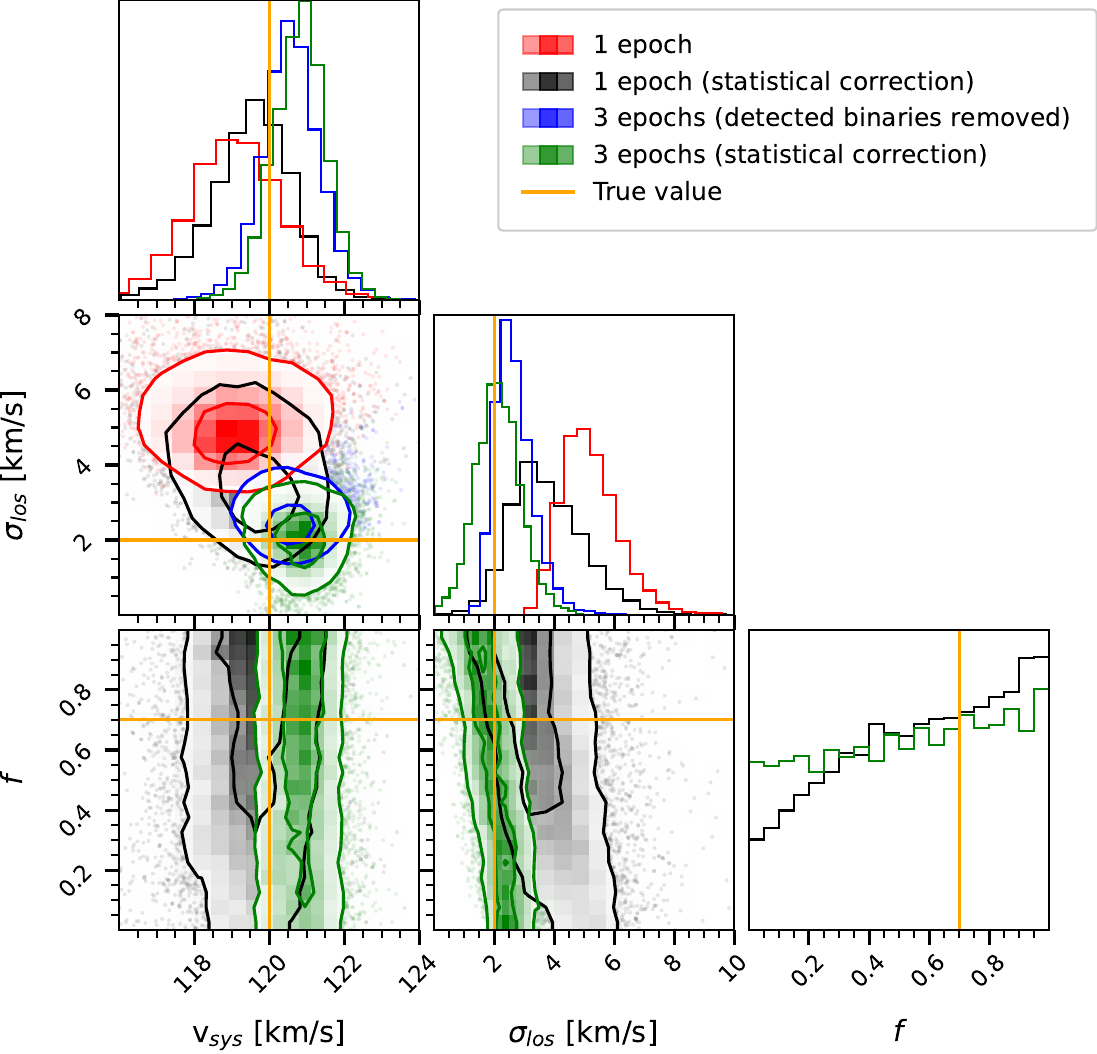}
            \caption{Cornerplot showing the PPD on $\sigma_{los}$, $v_{sys}$ and $f$ for a mock catalog of 20 stars in two cases. Single-epoch observations (red: not correcting for binaries; black: applying a statistical correction for undetected binaries), and 3-epochs observations (blue: by excluding stars with velocity variability $>3 \times$ $\Delta v_{los}$; green: as the blue, but also including a statistical correction for remaining undetected binaries). The contours indicate the 1- and 2-$\sigma$ ranges. The underlying values used to generate the mock are indicated with orange lines.}
        \label{fig:Fig2}
\end{figure*}

In Fig.~\ref{fig:Fig2} we show the inference on the $\sigma_{los}$, $v_{sys}$ and $f$ for this mock galaxy in four different cases. The red and black PPDs show the results when applying the same procedure as in Sec.~\ref{sec:results}., i.e., computing \slos from the l.o.s. velocities measured at epoch 0, neglecting the presence of binaries (red) or assuming a flat prior on $f$ (black). Neglecting the effect of binaries prevents an accurate inference of the true $\sigma_{los}$. In contrast, accounting for binaries significantly improves the results: the correct \slos can be marginally recovered, and $v_{sys}$ is also better reproduced. The blue and green PPDs show the result when analyzing the multi-epoch data, either by removing evident binaries but not accounting for undetected ones (blue), or by removing binaries while also accounting for undetected ones (green), following the method explained in this section. It is clear that the multi-epoch approach is crucial for correctly constraining $\sigma_{los}$. Even without accounting for the presence of remaining undetected binaries and just excluding stars with clear velocity variations (blue PPD), the intrinsic \slos is recovered with significantly better precision and accuracy compared to either of the single-epoch data cases. It is well-known that a multi-epoch strategy with a minimum time baseline of one year reduces significantly the inflation by binary stars \citep{Minor2010}. However, accounting for those binary stars that remain undetected still slightly affects the inference on $\sigma_{los}$. A dedicated set of mock experiments and a statistical analysis would be required to quantify the residual level of contamination after multi-epoch observations and to fully assess the performance of this methodology in such cases. We plan to carry out such an analysis in future work.

We emphasize that the test performed in this section is not intended to predict the general precision and accuracy achievable with a multi-epoch strategy. Its goal was to confirm that the mere availability of precise \vlos measurements at a few epochs reduces the inflation of $\sigma_{los}$, as already known from literature \citep{Minor2010}. This holds even when applied in a model-independent manner, i.e., by simply excluding stars with clear velocity variations. It also aims to show that correcting for undetected binaries after multi-epoch observations still affects the inference of $\sigma_{los}$; however, this step necessarily relies on assumptions about the underlying binary population, and its performance needs to be quantified through a more comprehensive analysis.

\section{Observed vs intrinsic \slos in the low-number statistics regime} \label{sec:obs_vs_int}

We have demonstrated that the effect of unresolved binaries on the observed velocity dispersion of UFDs is generally non-negligible, in agreement with previous studies \citep[e.g.,][]{McConnachie2010, Spencer2017,Gration2025}. In particular, \citet{Spencer2017} performed mock experiments predicting the observed \slos as a function of the intrinsic value and $f$ for systems containing 100 stars. The binary star models we adopt \citep{Moe2017} are similar to theirs \citep{Duquennoy1991}, and thus their results remain broadly consistent when the experiment is repeated within our framework. However, most UFDs rarely have as many as 100 member stars with reliable \vlos measurements (see col.~2 of Tab.~\ref{tab.1}). We therefore repeated the experiment under the more conservative assumption of only 10 stars, which substantially increases the uncertainties in the observed $\sigma_{los}$.

\begin{figure*}
    \centering
        \includegraphics[width=\textwidth]{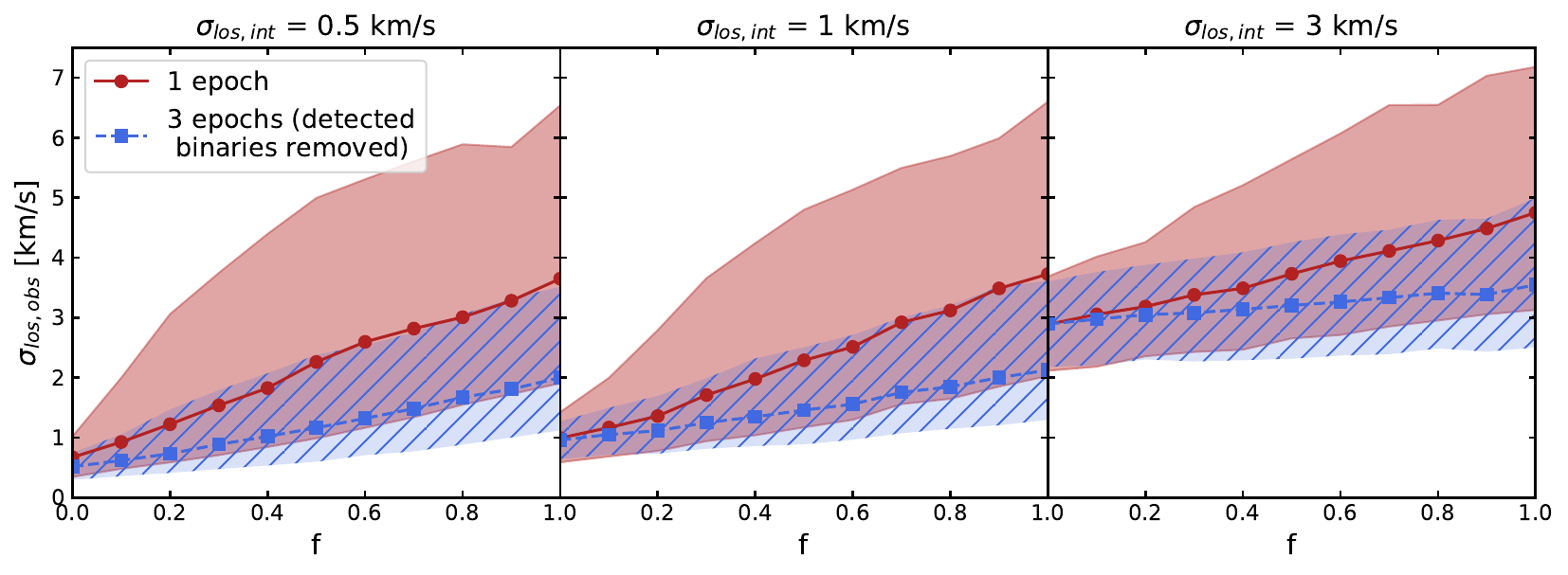}

        \caption{Observed velocity dispersion as a function of the underlying binary fraction for mock datasets of 10 stars with intrinsic \slos values of 0.5~\kms, 1~\kms, and 3~\kms. The red solid curves correspond to single-epoch observations without correction for the presence of binary stars, while the blue dashed curves refer to a dataset with three epochs and a one-year time baseline after excluding stars exhibiting velocity variability $>3 \times$  $\Delta v_{los}$. The lines indicate the median recovered values, and the shaded regions denote the 1$\sigma$ confidence intervals.}
        \label{fig:Fig3}
    \end{figure*}

We conducted the analysis for two observational setups: a single-epoch case and a multi-epoch case with three epochs: $t_0$, $t_1=t_0+\text{3 months}$, and $t_2=t_0+\text{12 months}$, resulting in a total time baseline of one year. In the multi-epoch case, we remove evident binaries and compute the error-weighted mean \vlos of each star for computing $\sigma_{los}$; this would be equivalent to the blue case of the previous section. We assume a fixed velocity error for the individual exposures of 1~\kms. The results for three mock systems with intrinsic \slos of 0.5~\kms, 1~\kms, and 3~\kms are presented in Fig.~\ref{fig:Fig3}. In the single-epoch case, the median recovered \slos values are in very good agreement with those reported by \citet{Spencer2017}, although we find a slightly lower inflation due to binaries for the same $f$. This difference likely arises from the stricter lower-period cutoff applied to the binary period distribution in our models. The main difference, however, is the width of the 1$\sigma$ range, which is significantly broader in our simulations due to the lower number of stars. Notably, for a binary fraction of 0.6, the observed \slos can reach values as high as 5.5~\kms within the 1$\sigma$ interval, regardless of the intrinsic $\sigma_{los}$. Consequently, all the dynamical masses and J-factors that one could compute with the observed \slos below this value and with low statistics are potentially inflated by binaries. In contrast, in the multi-epoch case, just by removing evident binaries but without any statistical correction, the upper limit for the 1-$\sigma$ range decreases to less than 3~\kms for the systems with lower intrinsic $\sigma_{los}$. This result underscores the importance of multi-epoch data when analyzing systems with a small number of kinematic tracers.

\cite{Gration2025} employ the same models as \cite{Spencer2017} and provide prescriptions for the median expected contamination due to binaries, assuming a binary fraction $f = 1$. However, they find that binaries produce an \slos of 7.3~\kms, significantly larger than our results. This discrepancy likely arises from the different lower-period cuts adopted. In our work, the minimum orbital distance, which depends on the mass ratio of the two stars, is always larger than the radius of an RGB star (0.21 au). In contrast, the models of \cite{Gration2025} include binary systems with separations as small as 0.001 au, as they treat the stars as point particles. Consequently, their models include more short-period binaries with large velocity amplitudes compared with ours.

\section{Conclusions} \label{sec:conclusions}

In this work, we have developed a flexible and self-consistent formalism to correct velocity dispersion measurements of UFDs for the effect of unresolved binary stars. By explicitly modeling the velocity distribution induced by binaries and incorporating it into a mixture-model framework, our approach avoids imposing priors on the binary fraction and can be applied to both single-epoch and multi-epoch spectroscopic datasets. We implemented tailored corrections for each observational regime and applied the single-epoch formalism to UFDs with published velocity dispersions below 4.5~\kms, as most of these systems currently only have single-epoch data, or for a few cases have very limited and often heterogeneous multi-epoch information.  We further tested the multi-epoch methodology on a simulated UFD mock composed of 20 stars observed at three different epochs over a baseline of two years.

We find that accounting for binary stars systematically decreases the estimated dynamical masses enclosed within $r_{1/2}$ by factors of 1.5–3 compared to estimates that do not account for binaries. These corrections have important astrophysical and cosmological implications, since the reduced inferred masses significantly weaken the prospects for detecting dark-matter decay or annihilation signals from UFDs, thereby increasing current upper limits on the relevant particle-physics cross-sections. At the same time, the lower masses estimated for systems such as Crater~II and Bootes I exacerbate existing tensions with $\Lambda$CDM predictions, as their mass-to-size ratio is difficult to reconcile with canonical NFW halos orbiting the Milky Way, as predicted by the $\Lambda$CDM paradigm.

Our results also call into question the classification of several systems—Leo~IV, Sag~II, and Uni~1—as bona fide galaxies. Once binary stars are accounted for, their velocity dispersions are no longer resolved, which combined with their unconstrained metallicity spreads \citep{Jenkins2021, Longeard2021, Simon2024} does not allow us to confirm whether these objects are genuine galaxies or globular clusters with the existing data. In contrast, systems such as Ret~II can be confidently confirmed as galaxies, as they display a clearly resolved \slosf even when binary stars are included, consistent with the metallicity spread measured in this system \citep{luna2025}.

After correcting for binaries, only Boötes~I and Crater~II show resolved velocity dispersions below 3~\kms, both of which benefit from relatively large kinematic samples ($\gtrsim$30 stars). For smaller samples, our analysis shows that confidently measuring velocity dispersion requires multi-epoch observations. The formalism presented here is specifically designed to handle such datasets, and we have demonstrated its ability to recover precise and unbiased \slos measurements with the proper multi-epoch data for a mock galaxy.

Finally, we present general predictions for the expected inflation of observed velocity dispersions caused by binaries in UFDs with intrinsically low dispersions (1–3~\kms), even when kinematic data are available for as few as ten stars. We find that, for a binary fraction of $f = 0.6$ and single-epoch data, the observed $\sigma_{\rm los}$ can reach 5.5~\kms or higher within the 1$\sigma$ interval, regardless of how low the intrinsic $\sigma_{\rm los}$ is. In contrast, with a one-year multi-epoch dataset, this value is significantly reduced to just 3~\kms for systems with low intrinsic $\sigma_{\rm los}$. These results highlight the critical importance of time-domain spectroscopy for establishing the dynamical nature of the faintest galaxies and for robustly inferring their DM content.

\begin{acknowledgements}
The authors acknowledge the referee for the constructive and detailed report, which enhanced the quality of the manuscript.

J. M. Arroyo acknowledges support from the Agencia Estatal de Investigación del Ministerio de Ciencia en Innovación (AEI-MICIN) and the European Social Fund (ESF+) under grant PRE2021-100638.

J. M. Arroyo, G. Battaglia and G. F. Thomas acknowledge support from the Agencia Estatal de Investigación del Ministerio de Ciencia, Innovación y Universidades (MCIU/AEI) under grant EN LA FRONTERA DE LA ARQUEOLOGÍA GALÁCTICA: EVOLUCIÓN DE LA MATERIA LUMINOSA Y OSCURA DE LA VÍA LÁCTEA Y LAS GALAXIAS ENANAS DEL GRUPO LOCAL EN LA ERA DE GAIA. (FOGALERA) and the European Regional Development Fund (ERDF) with reference PID2023-150319NB-C21 / 10.13039/501100011033  and PID2020-118778GB-I00 / 10.13039/501100011033

G. F. Thomas acknowledges the grant RYC2024-051016-I funded by MCIN/AEI/10.13039/501100011033 and by the European Social Fund Plus.
\end{acknowledgements}

\bibliographystyle{aa}
\bibliography{ref.bib}

\begin{appendix}
\onecolumn
\section{EMCEE specifics}\label{app:emcee}

\begin{figure*}
    \centering
        \includegraphics[width=0.93\textwidth]{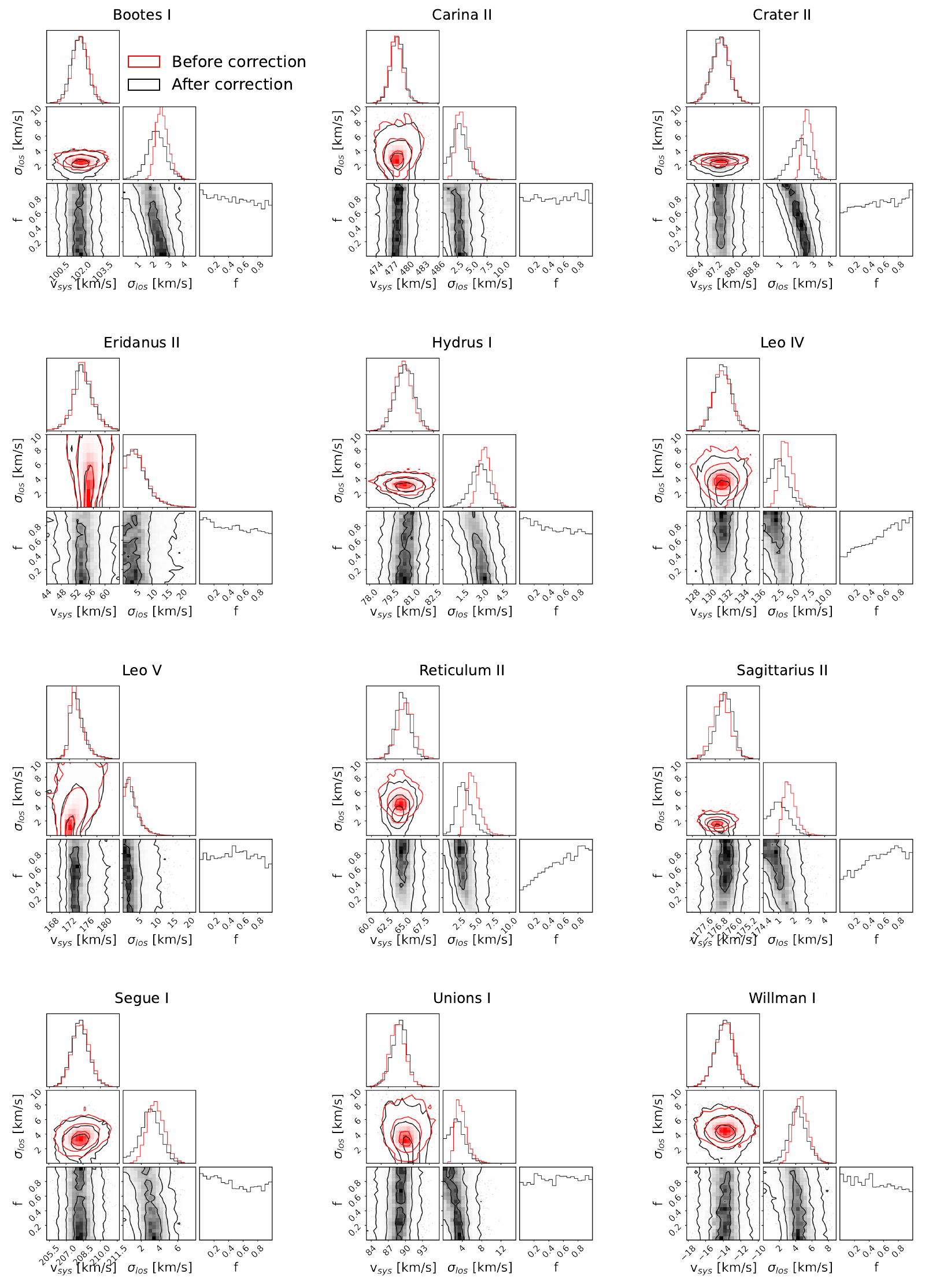}

        \caption{Inference on $\sigma_{los}$, $v_{sys}$ and $f$ for all the UFDs considered in this work, as indicated in the titles. The results of the analysis when not including the impact of undetected binaries are shown in red uncorrected by binaries, while those including the correction and assuming a uniform prior in f are illustrated in black, as indicated in the legend. } 
        \label{fig:FigB1}
    \end{figure*}

As priors, we always use uniform distribution between the limits we list in the following. For the runs not accounting for binary systems, i.e., $f$=0, we use the limits $v_{sys}  [kms^{-1}]= [-500, 500]$ and $\sigma_{los} [kms^{-1}] = [0, 25]$. After these runs, we use the median of the PPD $v_{sys}^{f=0}$ to set the priors for the runs in which we account for the presence of binary stars; the limits then are $v_{sys} = [v_{sys}^{f=0}-10, v_{sys}^{f=0}+10]$, $\sigma_{los} = [0, 25]$ and $f = [0, 1]$. For each run, we use 10 walkers, 4000 steps and set the burn-in to 500. We convolve $\mathcal{B}_i$ and $N_i(v_{sys}, \sigma_{los})$ in a grid of $-$50 to 50~\kms with a step of 0.01~\kms. We have checked that these hyperparameters are enough to systematically reach stability in the sampling of the PPD. The PPD of the parameters for the runs correcting by binary stars and with $f$=0 can be seen in Fig.~\ref{fig:FigB1} in black and red, respectively. In Tab.~\ref{tab.1} the median and uncertainties for the \slos in the different cases are presented.

\begin{sidewaystable*}
\centering
\caption{Observed l.o.s. velocity dispersions and dynamical masses of UFDs.}
\renewcommand{\arraystretch}{2}
\begin{tabular}{|c|c|c|c|c|c|c|c|c|c|c|c|c|}
\hline
Glx. & \# &$a_{1/2}$& \sloslit &\slosfz & \logMfz & \slosf & \logMf  & \slosfzs & \logMfzs  & Stars & $\log(J)$ & ref \\
& & [$arcmin$] & [\kms] & [\kms] & [$M_{\odot}$] & [\kms] & [$M_{\odot}$] & [\kms] & [$M_{\odot}$] &    & [$GeV^{2}cm^{-5}$]&\\
\hline
Boo~I           & 30 &  12.80$^{+0.70}_{-0.70}$& 2.4$^{+0.9}_{-0.5}$ & 2.55$^{+0.45(+1.70)}_{-0.37(-0.96)}$ & 6.08$^{+0.18(+0.48)}_{-0.18(-0.45)}$ & 2.18$^{+0.59(+1.91)}_{-0.58(-2.03)}$ & 5.95$^{+0.25(+0.58)}_{-0.31(-2.38)}$ & 1.97$^{+0.58(+1.88)}_{-0.54(-1.81)}$ & 5.86$^{+0.26(+0.62)}_{-0.32(-2.20)}$ & RGB & 16.14$^{+0.49(+1.17)}_{-0.62(-4.76)}$ & 1 \\
Car~II          & 14 &  8.69$^{+0.75}_{-0.75}$& 3.4$^{+1.2}_{-0.8}$ & 3.24$^{+1.30(+6.53)}_{-0.91(-2.07)}$ & 5.86$^{+0.34(+1.00)}_{-0.33(-0.93)}$ & 2.77$^{+1.43(+5.80)}_{-1.32(-2.72)}$ & 5.72$^{+0.41(+1.03)}_{-0.61(-3.65)}$ & 2.55$^{+1.40(+5.19)}_{-1.40(-2.52)}$ & 5.65$^{+0.43(+1.01)}_{-0.74(-3.88)}$ & RGB & 17.56$^{+0.82(+2.06)}_{-1.22(-7.30)}$ & 2 \\
Cra~II          & 60 &  31.20$^{+2.50}_{-2.50}$& 2.7$^{+0.3}_{-0.3}$ & 2.65$^{+0.32(+1.06)}_{-0.27(-0.69)}$ & 6.81$^{+0.15(+0.34)}_{-0.14(-0.31)}$ & 2.17$^{+0.47(+1.33)}_{-0.54(-1.68)}$ & 6.64$^{+0.22(+0.46)}_{-0.30(-1.33)}$ & 1.95$^{+0.41(+1.30)}_{-0.41(-1.12)}$ & 6.55$^{+0.21(+0.49)}_{-0.25(-0.80)}$ & RGB & 14.95$^{+0.44(+0.93)}_{-0.60(-2.67)}$ & 3 \\
Eri~III          & 8 &  0.32$^{+0.03}_{-0.04}$& 0.0$^{+9.1}_{-0.0}$\hyperlink{a}{[a]} & 4.83$^{+4.11(+18.28)}_{-3.01(-4.79)}$ & 5.20$^{+0.60(+1.43)}_{-0.94(-4.10)}$ & 4.45$^{+4.03(+17.84)}_{-2.91(-4.41)}$ & 5.12$^{+0.63(+1.47)}_{-1.01(-4.22)}$ & 4.63$^{+4.02(+16.85)}_{-3.34(-4.61)}$ & 5.16$^{+0.61(+1.40)}_{-1.20(-4.82)}$ & MS & 18.73$^{+1.26(+2.94)}_{-2.03(-8.45)}$ & 4 \\
Hyd~I           & 33 &  7.42$^{+0.62}_{-0.54}$& 2.7$^{+0.5}_{-0.5}$ & 3.17$^{+0.52(+2.09)}_{-0.45(-1.10)}$ & 5.70$^{+0.18(+0.48)}_{-0.17(-0.41)}$ & 2.77$^{+0.64(+2.01)}_{-0.69(-2.44)}$ & 5.58$^{+0.22(+0.52)}_{-0.29(-1.91)}$ & 2.55$^{+0.57(+1.93)}_{-0.65(-2.01)}$ & 5.51$^{+0.22(+0.53)}_{-0.30(-1.39)}$ & RGB & 18.08$^{+0.45(+1.03)}_{-0.58(-3.82)}$ & 5 \\
Leo~IV          & 20 &  2.61$^{+0.32}_{-0.32}$& 2.7$^{+1.2}_{-1.0}$ & 3.74$^{+1.20(+4.78)}_{-0.95(-2.52)}$ & 6.14$^{+0.31(+0.78)}_{-0.33(-1.04)}$ & 2.54$^{+1.50(+4.90)}_{-1.44(-2.52)}$ & 5.80$^{+0.47(+1.00)}_{-0.79(-4.15)}$ & 2.16$^{+1.49(+4.86)}_{-1.35(-2.14)}$ & 5.66$^{+0.52(+1.09)}_{-0.92(-4.40)}$ & RGB & 16.20$^{+0.93(+2.00)}_{-1.59(-8.30)}$ & 6 \\
Leo~V           & 10 &  1.05$^{+0.39}_{-0.39}$& 3.2$^{+1.7}_{-1.4}$ & 2.44$^{+2.47(+10.76)}_{-1.47(-2.41)}$ & 5.35$^{+0.76(+1.62)}_{-1.02(-4.05)}$ & 2.34$^{+2.54(+13.88)}_{-1.50(-2.32)}$ & 5.31$^{+0.79(+1.83)}_{-1.11(-4.28)}$ & 2.16$^{+2.69(+19.48)}_{-1.50(-2.14)}$ & 5.24$^{+0.86(+2.15)}_{-1.25(-4.43)}$ & RGB & 16.10$^{+1.58(+3.67)}_{-2.21(-8.56)}$& 6 \\
Ret~II          & 23 &  6.30$^{+0.40}_{-0.40}$& 3.2$^{+1.6}_{-0.5}$ & 4.53$^{+1.11(+4.52)}_{-0.83(-1.91)}$\hyperlink{b}{[b]} & 5.85$^{+0.24(+0.65)}_{-0.22(-0.52)}$ & 3.17$^{+1.27(+4.86)}_{-0.98(-2.90)}$ & 5.53$^{+0.34(+0.85)}_{-0.37(-2.21)}$ & 3.01$^{+1.22(+4.98)}_{-0.93(-2.57)}$ & 5.49$^{+0.34(+0.89)}_{-0.37(-1.71)}$ & RGB & 17.92$^{+0.68(+1.71)}_{-0.74(-4.42)}$ & 7 \\
Sag~II          & 19 &  1.70$^{+0.50}_{-0.50}$& 1.7$^{+0.5}_{-0.5}$ & 1.71$^{+0.46(+2.04)}_{-0.34(-0.77)}$ & 4.95$^{+0.36(+0.83)}_{-0.38(-0.71)}$ & 1.06$^{+0.65(+2.16)}_{-0.58(-1.04)}$ & 4.53$^{+0.57(+1.12)}_{-0.88(-3.87)}$ & 0.87$^{+0.65(+1.96)}_{-0.49(-0.85)}$ & 4.36$^{+0.63(+1.17)}_{-0.90(-3.35)}$ & RGB & 15.95$^{+1.14(+2.24)}_{-1.76(-7.74)}$ & 8 \\
Seg~1           & 64 &  3.93$^{+0.42}_{-0.42}$& 3.7$^{+1.4}_{-1.1}$ & 3.56$^{+0.99(+3.07)}_{-0.88(-2.48)}$ & 5.41$^{+0.30(+0.62)}_{-0.33(-1.13)}$ & 3.01$^{+1.07(+3.29)}_{-1.12(-2.97)}$ & 5.26$^{+0.35(+0.73)}_{-0.49(-3.70)}$ & 2.80$^{+1.03(+3.70)}_{-1.15(-2.75)}$ & 5.20$^{+0.36(+0.82)}_{-0.55(-3.59)}$ & RGB & 18.71$^{+0.69(+1.45)}_{-0.98(-7.39)}$ & 9 \\
Uni~1            & 12 &  0.90$^{+0.40}_{-0.30}$& 3.7$^{+1.4}_{-1.0}$ & 3.74$^{+1.41(+6.79)}_{-1.02(-2.43)}$ & 4.38$^{+0.48(+1.10)}_{-0.49(-1.13)}$ & 2.82$^{+1.70(+6.62)}_{-1.67(-2.80)}$ & 4.14$^{+0.61(+1.25)}_{-0.99(-4.47)}$ & 2.48$^{+1.86(+6.93)}_{-1.46(-2.44)}$ & 4.02$^{+0.69(+1.36)}_{-0.99(-3.81)}$ & MS  & 20.20$^{+1.22(+2.50)}_{-1.99(-8.94)}$ & 10 \\
Wil~1           & 40 &  2.52$^{+0.21}_{-0.21}$& 4.0$^{+0.8}_{-0.8}$ & 4.64$^{+0.95(+3.08)}_{-0.83(-2.25)}$ & 5.61$^{+0.28(+0.56)}_{-0.29(-0.70)}$ & 4.24$^{+1.10(+3.83)}_{-1.04(-3.98)}$ & 5.53$^{+0.32(+0.67)}_{-0.36(-2.54)}$ & 4.11$^{+1.04(+4.24)}_{-1.11(-3.95)}$ & 5.50$^{+0.31(+0.73)}_{-0.39(-2.92)}$ & RGB & 18.74$^{+0.63(+1.35)}_{-0.72(-5.09)}$ & 11 \\

\hline 

\end{tabular}
\tablefoot{Observed \slos and estimated masses of UFDs. Col.~1 lists the UFDs considered in this work. Col.~2 indicates the number of stars used to infer the \slos for each galaxy. Col.~3 shows the semi-major axis projected half-light radius. Col.~4 lists the values for \slos reported in literature. Col.~5 and Col.~6 indicate the observed \slos and the estimated mass using Eq.~\ref{eq:wolf} neglecting the effect of binaries. Col.~7 and Col.~ 8 the observed \slos and estimated mass assuming a flat prior on $f$. Col.~9 and Col.~10 list both parameters, assuming  $f=0.7$. Col.~11 shows whether the dominant component of the dataset is composed of RGB or MS stars. Col.~12 indicates the DM annihilation J-factor computed using $M(f)$. Finally, Col.~13 lists the references the \vlos data was taken from; the numbers indicate 1.~\cite{Koposov2011}, 2.~\cite{Li2018b}, 3.  \cite{caldwell2017}, 4.~\cite{Simon2024}, 5.~\cite{Koposov2018}, 6.~\cite{Jenkins2021}, 7.~\cite{Koposov2015}, 8.~\cite{Longeard2021}, 9.~\cite{Simon2011}, 10.~\cite{Smith2024} and 11.~\cite{Willman2011}. For $a_{1/2}$, ellipticity and modulus distance used to compute $R_{1/2}$ in physical units, the values for UNIONS I were taken from \cite{Simon2024}, while the rest are from \cite{Battaglia2022}. 

[\hypertarget{a}{a}] For Eridanus III the value shown for the literature is the upper value for a 90\% confidence. To see the comparison with this particular value, see App.~\ref{app:data}.

[\hypertarget{b}{b}] The reason why this value is different with respect to the literature can be found in App.~\ref{app:data}. In the same App. all the differences in the analysis with respect to original works are described for all the galaxies.} \label{tab.1}
\end{sidewaystable*}

\twocolumn
\section{Data} \label{app:data}
In this appendix, we discuss the specifics of the datasets used for each system. It is important to note that some works use multi-epoch data, while in the correction we perform, we assume that the data are single-epoch observations. However, in most of the cases, there are very few stars with time baselines longer than 1 month, and almost none with time baselines longer than 1 year. We checked that, in this case, the correction for undetected binaries is very similar to the one we perform. Unless said otherwise, we use the photometric parameters ($L$ and $a_{1/2}$) listed in \cite{battaglia2022b}, where the original references can be found. Hereafter, we refer to the velocity dispersion values obtained in the literature as \sloslit and to those obtained in this work using $f=0$ as $\sigma_{los, f\text{=0}}$.

Boo~I: We use the data presented in \cite{Koposov2011}, selecting from Table 1 those members with "Best flag" = 'B'. The velocities are the combined average of 16 repeated measurements obtained over a maximum baseline of nearly one month. Due to the multi-epoch information, they identified 1 RR-Lyrae star that we removed from the dataset. The total time baseline is too short and the individual uncertainties too large for making the dataset suitable for multi-epoch analysis of binaries. The authors found the velocity distribution of Boo~I to be composed of two different components, a "cold" one with \sloslit = $2.4^{+0.9}_{-0.5}$~\kms and a hot one with \sloslit = $4.6^{+0.8}_{-0.6}$~\kms. We only consider stars belonging to the "cold component" by applying a 3 \sloslit(cold component)-clipping to the \vlos distribution. The small differences in the inferred values of \slos are due to the authors fitting the \vlos distribution with the sum of two Gaussians rather than performing a sigma-clipping.

Car~II: We use the data presented in \cite{Li2018b}, which consists of 14 stars: 8 RGB and 6 blue horizontal branch, plus two RR Lyrae stars that we removed from the dataset. This dataset contains multi-epoch data collected with different telescopes. However, the maximum baseline is 5 months for 2 stars, less than 1 month for 1 star, and only a couple of days for 2 stars. We removed 2 binary stars the authors detected. Only 3 stars out of 14 were observed for 1 month or more, so this dataset is not suited for the multi-epoch analysis. The very small differences between \sloslit and \slosfz could be due to minor differences in the averaging process of individual velocities or in the priors in the EMCEE runs. 

Cra~II: We used the dataset from \cite{caldwell2017} with no multi-epoch information. No membership information was provided. Therefore, we cross-matched the sample with the catalog from \cite{Battaglia2022} and selected all the stars with probabilities of membership greater than 0.07 as members. Afterwards, we removed 4 stars with inconsistent \vlos by applying a 3-\sloslit clipping, resulting in a final total of 60 reliable velocity members. According to the selection in \cite{caldwell2017}, there are 62 reliable velocity members, and the \slos value they report coincides very well with ours.

Eri~III: We use the data from \cite{Simon2024}. We take the velocities from the individual exposures and combine them using an error-weighted mean. They only have precise multi-epoch measurements for 1 star, for which they did not find velocity variation. The authors only placed 90\% and 95.5\% upper limits on $\sigma_{los}$, which are 9.1 and 10.8~\kms, respectively. If we compute those values using their priors ([0.19, 13.6]~\kms for $\sigma_{los}$), we infer 9.5 and 11.1~\kms; which is in excellent agreement with their inference. The photometric parameters we use are taken from \cite{koposov2015a}.

Hyd~I: We use the data from the single-epoch measurements of \cite{Koposov2018}. We select as members those stars with probabilities $10^{\text{'logodds'}}$ > 0.95 of being a Hyd~I member star versus being a foreground contaminant. There are some differences in the inferred \slos that can arise because of the different methodologies; the authors use a probabilistic approach including members and contaminants in a mixture model, while we use only the member stars. The photometric parameters are from \cite{Koposov2018}.

Leo~IV and Leo~V: We use the data exactly as presented in \cite{Jenkins2021} and select all the stars classified as members according to the subjective 'Member flag'. This dataset contains multi-epoch data with a total time baseline of 7/8 months for Leo~V/Leo~IV for a significant number of stars. Although the uncertainties are large, this is the best opportunity to apply the multi-epoch binary correction presented in this work. This could affect the inference for Leo~IV, but not for Leo~V, since the correction by binaries would not significantly impact the results, as \slosfz is already unconstrained. However, we could not apply the multi-epoch methodology because the velocities for the individual exposures were not provided. The authors found no binary stars in Leo~IV and two potential binaries in Leo~V. For Leo~V, we removed the star with ID Leo5-1034 as it shows clear velocity variation. However, we did not remove the star with ID Leo5-1038 as according to the reported p-value it is not a clear binary. Note that the authors provide different values of \slos for different prior selections. We quote the results of the model M2 as it is the one in which the authors use the same priors and selection of member stars as we do. 

Ret~II: We used the data from \cite{Koposov2015}. This dataset does not contain multi-epoch data. We selected all the stars with 'yes' or 'yes?' in the membership flag. The \slos we find is slightly different because the authors simultaneously model the background and the galaxy with a mixture model, while we use only the member stars.

Sag~II: We used the data from \cite{Longeard2021}. This dataset contains some multi-epoch data with a maximum time baseline of 1 month; no binaries were detected within the members. The methodologies are similar, and therefore our results are in very good agreement. We use the photometric parameters from \cite{Longeard2020}.

Uni~1: We used the data from \cite{Smith2024}. This dataset does not contain multi-epoch data. The slight difference between our inference of \slos and the one quoted by the authors may be due to the prior range in the EMCEE runs. We use the photometric parameters from \cite{Smith2024}.

Seg~1: We used the single exposures from \cite{Simon2011}. Although there is multi-epoch data with a total time baseline of one year, it is only for 20 stars out of the total 64 stars. Furthermore, the mean velocity error is around 6~\kms, so it is not properly suitable for a full multi-epoch analysis. From the individual exposures, we remove stars with velocity variations between epochs larger than three times the mean error of the observations. Then, we compute the error-weighted mean of the l.o.s. velocity for each individual star. Finally, we remove the individual star with ID 'J100704.35+160459.4', which significantly increases $\sigma_{los}$.

Wil~1: We used the data from \cite{Willman2011}. This dataset does not contain multi-epoch data. We removed all stars classified as likely non-members by the calcium triplet equivalent width < 2.3 \r{A} as the authors suggested, and used the same sample of 40 member stars they used. We obtained a slightly different \slos value compared to theirs, which could be due to slight differences in methodology. 

We cross-matched all the stars with the catalog of probabilities of membership from \cite{Battaglia2022}. All the stars in common, except for the star with ID 10 in Hyd~I \citep{Koposov2018}, have a probability of membership greater than 0.1, meaning that they are not certainly contaminants. Furthermore, the potential contaminant star from Hyd~I has an observed l.o.s. velocity that is consistent with the systemic velocity of the UFD, so we decided to keep it in the sample.

\section{Classical dwarf galaxies} \label{app:classical}

In this section, we test whether undetected binaries can alter the DM density profile inferred for classical dwarf spheroidal galaxies. We use mock galaxies generated with AGAMA \citep{Vasiliev2019}, emulating the Sculptor-like models from \cite{Arroyo-Polonio2025}, including both the metal-rich and metal-poor populations. Three different mocks are considered: a cored DM halo mock with a logarithmic inner slope $\gamma$ = 0 for the DM density profile, a cuspy DM halo mock ($\gamma =1$) and an intermediate one ($\gamma=0.5$). For each mock, we generate a corresponding “binary-contaminated” mock galaxy. To do this, we assume a binary fraction $f = 0.7$, use the binary star models presented in this work, and add the line-of-sight velocity component $v_{los,bin}$ for a single epoch of the binary stars. The mocks are then analyzed following the procedure described in \cite{Arroyo-Polonio2025}, neglecting the effect of binaries in all cases. This allows us to compare the inferred properties of a mock galaxy without binaries to those of a galaxy containing binary stars when their effect is ignored.

\begin{figure}[H]
    \centering
        \includegraphics[width=0.5\textwidth]{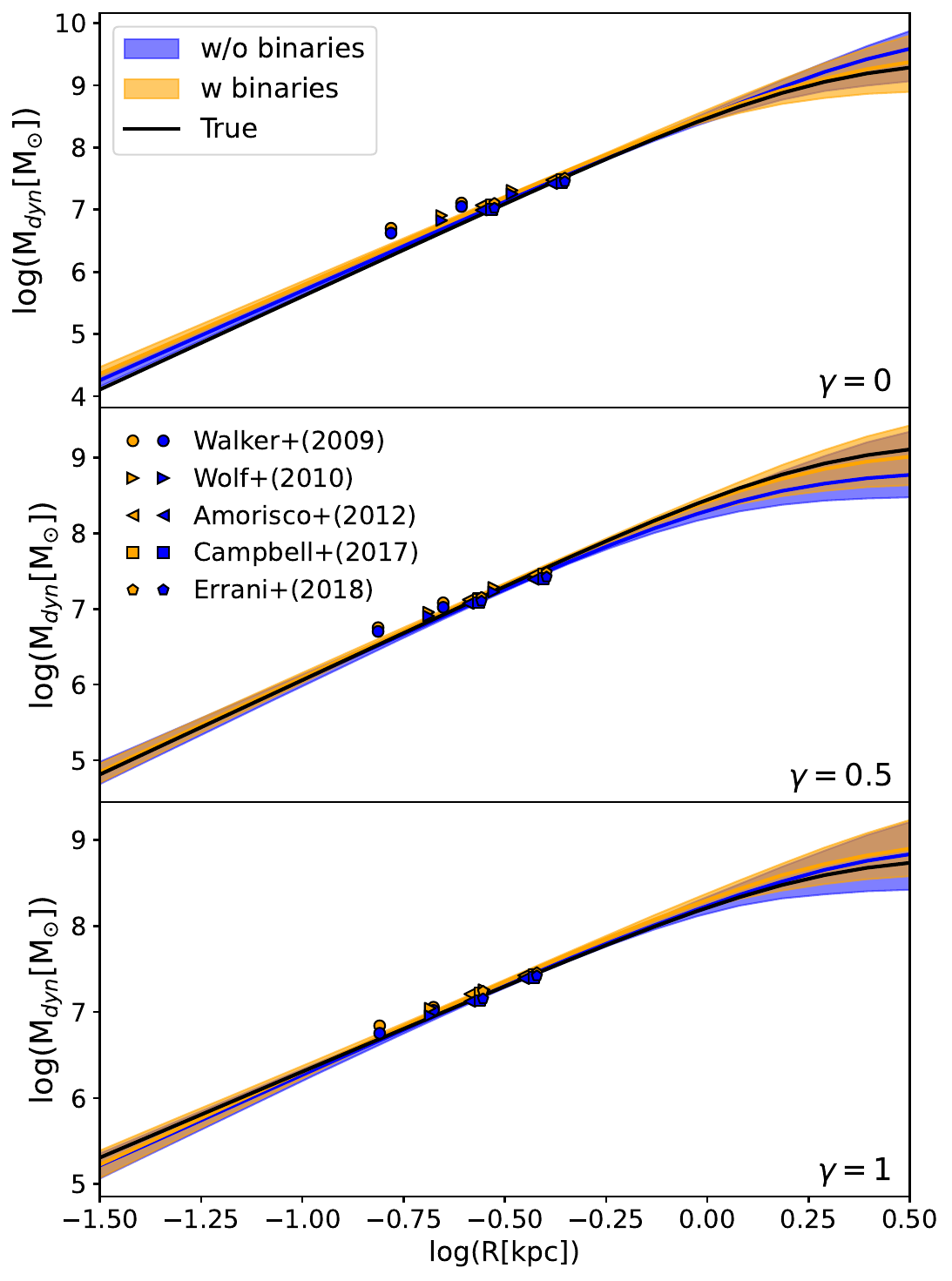}
    
        \caption{DM enclosed mass profiles for the three mock galaxies. From top to bottom they have $\gamma$ = 0, 0.5, and 1. The black line shows the true DM density profile used to generate the mock; the blue line and band show the median and 1-$\sigma$ range for the output parameters of the modeling of the underlying mock; the orange line shows the same for the modeling of the mock contaminated by binaries. The symbols indicate the value estimated by mass estimators, applied to both populations, with the same color-coding.} 
        \label{fig:FigC1}
    \end{figure}

The results are shown in Fig.~\ref{fig:FigC1}, alongside predictions from different mass estimators \citep{Walker2009a,Wolf2010, Amorisco2012, Campbell2017, Errani2018} applied to the individual stellar populations of these mocks. Blue lines indicate the mocks without binaries, while orange lines represent the binary-contaminated mocks. Although binaries slightly inflate the inferred enclosed mass profiles, the effect is minor for classical dwarf galaxies, or at least for Sculptor-like systems. The deviations are smaller than the typical uncertainties, suggesting that binaries do not need to be explicitly accounted for in dynamical modeling of these systems. In \cite{Wang2023}, the authors performed a similar analysis using simulated galaxies to investigate the effect of binaries on UFDs. They contaminated the galaxies with binary stars, but rescaled the line-of-sight velocity component of the binaries so that the ratio between the intrinsic \slos of the galaxy and the expected amplitude of the binary stars $v_{los,bin}$ matches the values expected in UFDs. In this case, binaries are found to have a more significant impact on the inferred mass profiles.

\section{Special cases: Aquarius III and Delve 1} \label{app:special}
Aquarius~III and Delve~1 were not considered in this work. The available datasets for these systems exhibit an observed velocity spread lower than the measurement errors reported in the literature. This may be due to an overestimation of the \vlos uncertainties or simply a statistical fluctuation due to small number statistics. In such cases, models that incorporate the presence of binary stars fail to reproduce the observed velocity distribution correctly, even when exploring an intrinsic \slos of zero. As a result, the posterior probability distribution becomes increasingly noisy as the binary fraction $f$ increases. Consequently, the inferred \slos becomes highly sensitive to the choice of prior, and larger values of $f$ can artificially produce higher inferred intrinsic dispersions, rendering the modeling unreliable.

\end{appendix}

\end{document}